\documentclass[useAMS,usenatbib]{mn2e}
\usepackage{graphicx}
\usepackage{mathrsfs}
\usepackage{commath}
\usepackage{amssymb}
\usepackage{float}
\usepackage{txfonts}
\bibpunct{(}{)}{;}{a}{,}{,}
\title[The $\sigma_s$--$\mathscr{M}$ relation in supersonic turbulence: I. Isothermal, magnetised gas]{The Density Variance--Mach Number Relation in Supersonic Turbulence: I. Isothermal, magnetised gas}
\author[Molina et al.]{F. Z. Molina$^{1}\thanks{Member of the International Max Planck Research School for Astronomy and Cosmic Physics at the University of Heidelberg (IMPRS-HD) and the Heidelberg Graduate School of Fundamental Physics (HGSFP)}$\thanks{E-mail:
molina@stud.uni-heidelberg.de}, S. C. O. Glover$^{1}$, C. Federrath$^{3,2,1}$, \& R. S. Klessen$^{1}$ \\
$^{1}$Zentrum f\"ur Astronomie der Universit\"at Heidelberg, Institut f\"ur Theoretische Astrophysik, Albert-Ueberle-Str. 2, 69120. Heidelberg, Germany\\
$^{2}$Ecole Normale Sup\'erieure de Lyon, Centre de Recherche Astrophysique. 46 All\'ee d'Italie, F-69364, France \\
$^{3}$Monash Centre for Astrophysics (MoCA), School of Mathematical Sciences, Vic 3800, Australia}
\begin{document}
\bibliographystyle{mn2e}
\date{Accepted: 2012 April 05. Received: 2012 April 05; in original form: 2011 November 07}

\pagerange{\pageref{firstpage}--\pageref{lastpage}} \pubyear{2012}

\maketitle

\label{firstpage}

\begin{abstract}
It is widely accepted that supersonic, magnetised turbulence plays a fundamental role for star formation in molecular clouds. It produces the initial dense gas seeds out of which new stars can form. However, the exact relation between gas compression, turbulent Mach number, and magnetic field strength is still poorly understood. Here, we introduce and test an analytical prediction for the relation between the density variance and the root-mean-square Mach number $\mathscr{M}$ in supersonic, isothermal, magnetised turbulent flows. We approximate the density and velocity structure of the interstellar medium as a superposition of shock waves. We obtain the density contrast considering the momentum equation for a single magnetised shock and extrapolate this result to the entire cloud. Depending on the field geometry, we then make three different assumptions based on observational and theoretical constraints: $B$ independent of $\rho$, $B\propto \rho^{1/2}$ and $B\propto \rho$. We test the analytically derived density variance--Mach number relation with numerical simulations, and find that for $B\propto \rho^{1/2}$, the variance in the logarithmic density contrast, $\sigma_{\ln \rho/\rho_0}^2=\ln[1+b^2\mathscr{M}^2\beta_0/(\beta_0+1)]$, fits very well to simulated data with turbulent forcing parameter $b=0.4$, when the gas is super-Alfv\'enic. However, this result breaks down when the turbulence becomes trans-Alfv\'enic or sub-Alfv\'enic, because in this regime the turbulence becomes highly anisotropic. Our density variance--Mach number relations simplify to the purely hydrodynamic relation as the ratio of thermal to magnetic pressure $\beta_0\to\infty$.
 
\end{abstract}

\begin{keywords}
ISM: structure -- clouds -- magnetohydrodynamics (MHD) -- shock waves -- stars: formation -- turbulence \\
\end{keywords}

\section{Introduction}
Detailed knowledge about the statistical characteristics of the density structure is of pivotal importance for many fields in astronomy and astrophysics. Probability distribution functions (PDFs) of the density have been introduced as a simple and robust measure of the one-point statistics for many applications, ranging from cosmology, where the Press-Schechter formalism was primarily established \citep{1974PressSchechter}, to star formation and theories of the initial mass function or the core mass function (e.g., \citealp*{1982Fleck,1984Zinnecker,1997PNJ,2000KlessenBurkert}; \citealp{2004Lietal}; \citealp*{2008HennebelleChabrier,2009HennebelleChabrier,2011PN}).

In the star formation context, the relation between the width of the density PDF -- the density variance or standard deviation -- and the root-mean-square (rms) Mach number in supersonic turbulent flow is a key ingredient for analytical models of the star formation rate \citep{2005KrumholzMckee,2011PN}, and for the stellar initial mass function or the core mass function \citep{2002PN,2008HennebelleChabrier,2009HennebelleChabrier}. In this framework, supersonic turbulence plays a fundamental role in determining the density and velocity statistics of the interstellar medium \citep{2004Elmegreen,2007McKeeOstriker} and controls stellar birth \citep{2004MacLowKlessen}. Conversely, the importance of magnetic fields in the star formation process is still inconclusive, despite decades of research (\citealp{1999MouschoviasCiolek,2007McKeeOstriker}; \citealp*{2009Crutcheretal}; \citealp{2010Crutcheretal,2012Bertrametal}). Hence, the question of how magnetic fields affect the density variance--Mach number relation is still not clearly answered, despite the empirical findings of \citet{2001Ostrikeretal} and the analytical ansatz provided by \citet{2011PN}. 

For purely hydrodynamical, supersonic, isothermal, turbulent gas, the relation between the density variance and Mach number has been identified and widely studied in numerical simulations (e.g., \citealp{1997PNJ,1998PassotyVS}; \citealp{2008Federrathetalb}; \citealp*{2008Federrathetala}; \citealp{2010Federrathetal}; \citealp*{2011Priceetal}). This relation is commonly assumed to be linear,

\begin{equation}\label{linear-s}
\sigma_{\rho/\rho_0}=b\mathscr{M},
\end{equation}
where $\sigma_{\rho/\rho_0}^2$ is the density variance (to emphasise the density fluctuations about the mean $\rho_0$, it makes sense to express the density in terms of the density contrast $\rho/\rho_0$), $b$ is a proportionality constant of order unity as explained in more detail below, and $\mathscr{M}$ is the rms Mach number.  Usually, the density contrast is written in terms of its logarithm, $s\equiv \ln(\rho/\rho_0)$.

Several authors have noted that the PDF of the logarithm of the density contrast $s$ -- produced by supersonic turbulent flow of isothermal gas -- follows approximately a lognormal distribution (e.g. \citealp{1994Vazquez-Semadeni, 1997PNJ, 1998PassotyVS,1999NP, 2000Klessen}; \citealp*{2001Ostrikeretal}; \citealp{2003Lietal,2007Kritsuketal,2008Federrathetala, 2008Lemaster, 2009Schmidtetal,2010Gloveretal, 2010Federrathetal, 2011PN}; \citealp{2011Collinsetal, 2011Priceetal}),

\begin{equation}\label{log-norm}
p_s \dif s=\frac{1}{\sqrt{2\pi\sigma_s^2}}\exp \left[-\frac{\left(s- s_0\right)^2}{2\sigma_s^2}\right] \dif s,
\end{equation}
where the mean $ s_0$ is related to the density variance by $s_0 = -\sigma_s^2/2$, due to the constraint of mass conservation. Besides the empirical findings of \citet{1994Vazquez-Semadeni}, \citet{1997PNJ}, and \citet{1998PassotyVS}, there is no clear explanation for the shape of the PDF. From a mathematical point of view, a log-normal distribution is the result of independent random perturbations driven in a stationary system \citep{1993PopeChing} as a consequence of the central limit theorem \citep{1994Vazquez-Semadeni,1997PNJ,1999NP,2010Federrathetal}. The physical interpretation is that density fluctuations present at a given location are produced by successive passages of shocks with amplitudes independent of the local density. For a log-normal distribution, the density variance -- given by Equation~(\ref{linear-s}) -- is equivalent to

\begin{equation}\label{HD}
\sigma_s^2=\ln\left[1+{b^2\mathscr{M}^2}\right].
\end{equation}

The parameter $b$ in Equations~(\ref{linear-s}) and~(\ref{HD}) is related to the kinetic energy injection mechanism -- the forcing $\mathbf{F}$, which drives the turbulence. \citet{2008Federrathetala} found that $b=1$ for purely compressive (curl-free) forcing, $\nabla\times\mathbf{F}=0$, while $b=1/3$ for purely solenoidal (divergence-free) forcing, $\nabla\cdot\mathbf{F}=0$. In a follow-up study, \citet{2010Federrathetal} showed that $b$ increases smoothly from $1/3$ to $1$, when the amount of compressive modes, $F_\mathrm{comp}/(F_\mathrm{sol}+F_\mathrm{comp})$ is gradually increased from $0$ to $1$. For the natural mixture of modes, $F_\mathrm{comp}/(F_\mathrm{sol}+F_\mathrm{comp})=1/3$, which is also the mixture of forcing modes used in all our numerical experiments here, they found $b\approx0.4$, so we will later use that value for comparing our analytic model with numerical simulations.

When magnetic fields are included, the density variance is significantly lower than in the unmagnetised case for simulations with Mach numbers $\mathscr{M}\gtrsim 10$ \citep{2001Ostrikeretal,2011Priceetal}. Recently, \citet{2011PN} provided an analytical ansatz for the hydrodynamical density contrast in supersonic, turbulent flow, which in turn follows the approach of \citet{1980DysonWilliams} for obtaining the density contrast for strong adiabatic shocks, but extended to the magnetic case. Their $\sigma_s$--$\mathscr{M}$ relation was, however, not tested with numerical simulations.

The density PDF may or may not deviate from a log-normal form when other processes -- like heat exchange and gravitation -- are included. For example, when a non-isothermal equation of state is considered, the PDF still closely follows a log-normal distribution over a range of densities \citep[see e.g.,][]{2007GloveryMacLow}. However, depending on whether the equation of state is softer or harder than isothermal, it might acquire power-law tails either at high or low densities (\citealp{1998PassotyVS,1998Scaloetal,2001Wada}; \citealp*{2003Lietal,2007McKeeOstriker}). The density PDF also deviates from log-normal when gravity is included. In this instance, the PDF exhibits a power-law tail at high densities (\citealp{2000Klessen,2008Federrathetalb,2009Kainulainenetal,2011ChoKim}; \citealp*{2011Kritsuketal}). In addition, turbulent intermittency also leads to deviations from the log-normal PDF in the wings of the distribution \citep{2010Federrathetal}. Consequently, the accuracy of the measurement of the density variance, using Equation (\ref{log-norm}), may be compromised depending on the importance of the different processes involved in real molecular clouds.

Here, we present an analytical derivation for the $\sigma_s$--$\mathscr{M}$ relation in supersonic turbulent isothermal gas including magnetic fields. Our results are in qualitative agreement with \citet{2001Ostrikeretal} and \citet{2011Priceetal}, however, here we present quantitative predictions and tests. The present work is organised as follows: In \S\ref{ana-approach} we describe the analytical approach made for the $\sigma_s$--$\mathscr{M}$ relation. In this section, we start with the study of the density contrast of a single shock confined into a cubic box, and then we extrapolate it to the whole cloud in \S\ref{rhocontrast}. In \S\ref{sigma-s} we propose three $\sigma_s$--$\mathscr{M}$ relations given by three different assumptions of the behaviour of magnetic fields with density. We test these predictions with numerical simulations in \S\ref{test}, and conclude in \S4.

\section{Analytical derivation}\label{ana-approach}

Our basis for obtaining the density variance--Mach number relationship involves determining how the density contrast changes with the Mach number. The density variance $\sigma_{\rho/\rho_0}$ and the density contrast are related by:

\begin{equation}\label{std}
\sigma_{\rho/\rho_0}^2={1\over V}\int_V \left({\rho \over \rho_0}-1\right)^2\dif V,
\end{equation}
where $\rho$ is the local density, $\rho_0$ is the mean density in the volume, and $V$ is the volume of the cloud. The density contrast is a measure of the density fluctuations in the flow, and therefore it is useful for identifying the disturbances that originate from shock fronts and compressions.

\subsection{Density contrast in magnetohydrodynamics}\label{rhocontrast}

Supersonic turbulence in the interstellar medium generates a complex network of shock waves (or simply shocks). When the velocity of the fluid exceeds that of sound, it leads to the formation of shocks that are one of the most important distinctive effects of the compressibility of the fluid \citep[e.g.,][]{1987LandauLifshitz}. 

In order to study the density contrast in a molecular cloud, we first consider the physics of the discontinuity formed by a single shock front. We then generalise the results to the ensemble of shocks confined in a cloud. Following \citet{2005Lequeux}, we describe a shock by choosing two control surfaces, one on either side of the discontinuity, and parallel to each other. Let us choose the shock surfaces as the reference frame, such that the control surfaces are stationary with respect to the shock. We also define the ``parallel'' direction as the one parallel to the flow of gas through the shock (i.e., perpendicular to the shock front). From the well known equations of fluid dynamics, it is then possible to derive equations that expresses the conservation of matter and momentum flux for a magnetised inviscid, neutral fluid:

\begin{equation}\label{masseq}
v_{\parallel,1}\rho_1=v_{\parallel,2}\rho_2,
\end{equation}
and
\begin{equation}\label{origin}
\rho_1 \left(v_{\parallel,1}^2 + {c_{s,1}^2\over \gamma_1} + {v_{A{\mathbf \perp},1}^2\over 2} \right) =  \rho_2\left(v_{\parallel,2}^2+ {c_{s,2}^2\over \gamma_2} + {v_{A\perp,2}^2\over 2}\right), 
\end{equation}
respectively. In these equations, the subscripts 1 and 2 indicate the pre--and post-shock conditions, respectively.  The velocity of the gas into the shock is $v_\parallel$, while $c_s$ is the adiabatic sound speed, $\gamma$ is the ratio between the specific heats and $v_{A\perp}$ is the Alfv\'en velocity, defined here as $v_{A\perp}= B_\perp/(4\pi \rho)^{1/2}$, where $B_\perp$ is the magnetic field perpendicular to the flow direction. The post-shock density is described by $\rho_2$.

We now make two important approximations. First, as we wish to focus on the role of magnetic fields in determining the density variance, we assume that the gas is isothermal, deferring consideration of non-isothermal effects to future work. Our assumption of isothermality implies that $c_{s,1}=c_{s,2}=c_s$ and $\gamma_1=\gamma_2=1$. Second, as we are considering an entire molecular cloud, we approximate it as an ensemble of shocks. We assume that we can express the average pre-shock velocity in terms of the rms velocity $v_0$ -- hereafter, the subscript ``0'' indicates the volume averages -- as $v_{\parallel,0}^2=b^2v_0^2$, where the factor $b$ depends on the number of degrees of freedom available for the compressive modes \citep{2008Federrathetala}. We also assume that the typical pre-shock magnetic and thermal pressures are just those given by volume averages over the total volume, allowing us to write them in terms of the volume-averaged density $\rho_{0}$ and the rms Alf\'enic velocity $v_{A,0}$. Similarly, we assume that the typical pre-shock density is simply the volume-averaged density. Making these assumptions, and introducing the ratio of the thermal pressure to magnetic pressure
\begin{equation}\label{betafin}
\beta\equiv{P_{\rm{th}} \over P_{\rm{mag}}}={2}{c_{s}^2\over v_{A}^2},
\end{equation}
we can rewrite Equation (\ref{origin}) as

\begin{equation}\label{2-origin}
b^2\mathscr{M}^2{\rho_0\over \rho_2}\left(1-{\rho_0\over \rho_2}\right) +{\rho_0\over \rho_2}\left(1 + \beta_0^{-1}\right) = \left( 1 + \beta_2^{-1}\right),
\end{equation}
where the rms Mach number is given by $\mathscr{M}=v_0/c_{s}$. 

In order to solve this equation for the characteristic density contrast associated with the shocked gas, $\rho_{2} / \rho_{0}$, it is necessary to determine $\beta_{2}$, the post-shock ratio of the thermal to magnetic pressures. The value of this will depend on the change in the magnetic field strength through the shock, which in turn depends on the orientation of the field with respect to the flow of gas through the shock. Using magnetic flux and mass conservation during compression, one can show that $B\propto\rho^\alpha$ with $0\leq\alpha\leq1$, depending on the field geometry and direction of compression. In the extreme case where the gas flows in a direction parallel to the field lines, the field strength will be the same on either side of the shock despite the jump in density, and the field strength then will be independent of density, i.e., $\alpha=0$. In the other extreme case where the field is oriented at right-angles to the gas flow, the shock jump conditions for magnetic flux freezing imply that $B \propto \rho$, i.e., $\alpha=1$. Meanwhile, compression of an isotropic field along all three spatial directions gives $B\propto\rho^{2/3}$. However, for our ``average shock'', we expect behaviour that lies somewhere between $0\lesssim \alpha \lesssim 1$. By looking at observations and existing simulations, we can get some guidance as to what this intermediate behaviour should be.

Observationally, \citet{1999Crutcher} presented a study of the magnetic field strength in molecular clouds measured with the Zeeman effect. He fitted the results with a power law $B\propto \rho^\alpha$ and found that $\alpha=0.47 \pm 0.08$. \citet*{2003Crutcheretal} provided additional support for this result. More recently, \citet{2010Crutcheretal} have presented a detailed compilation of Zeeman data based on a much larger number of measurements. They find that at number densities 
$n<300$ cm$^{-3}$, the data is consistent with a field strength that is independent of density, while at higher densities they obtain  $B\propto \rho^{0.65\pm 0.05}$.

From a theoretical point of view, \citet{1999PN} noted that their $B$ distributions closely match the observational scaling given by \citet{1999Crutcher} and \citet{2003Crutcheretal}, $B\propto \rho^{1/2}$, for high $B$ in their high Alfv\'enic Mach number regime. \citet*{2001Kimetal} also study the relationship between $B$ and $\rho$, and find that $\alpha \simeq 0.4$, albeit with large scatter, especially at low densities. Additionally, \citet{2009Banerjeeetal} report that the magnetic field strength appears to scale in their simulations as $B\propto \rho^{1/2}$ for number densities $10^2\lesssim n\lesssim 10^4$ cm$^{-3}$, although with significant scatter around this value. On the other hand, \citet{2000HennebellePerault} found that the magnetic field does not necessarily increase with the density. Asides from these reports, if the magnetic flux is not conserved, but increases due to turbulent dynamo amplification during compression, $\alpha$ can become larger than the values quoted above, depending on the Reynolds numbers of the gas \citep{2010Schleicheretal,2010Suretal,2011Federrathetal}. Thus, even if the gas is compressed only parallel to the field lines, turbulent tangling of the field can lead to $\alpha>0$ during compression.

Given the different possible relations between the magnetic field strength and the density, we consider three cases to include in Equation (\ref{2-origin}): the two extreme cases, where $B$ is independent of the density, and where $B \propto \rho$, and an intermediate case with $B \propto \rho^{1/2}$. We also note that if we were to take instead the relation  $B\propto \rho^{0.65}$ suggested by the most recent observational data, then we would obtain results quite similar to the $B \propto \rho^{1/2}$ case.

\subsubsection{First case: $B$ independent of $\rho$}\label{1case}

We start by considering one extreme, the case where $B$ is independent of the density. In this scenario, Equation (\ref{2-origin}) becomes a second-order equation, independent of the magnetic field strength

\begin{equation*}
\left({\rho_2 \over \rho_0}\right)^2-\left(b^2\mathscr{M}^2+1\right)\left({\rho_2\over \rho_0}\right)+b^2\mathscr{M}^2=0.
\end{equation*}
This equation results in a density contrast

\begin{equation}\label{conte}
{\rho_2 \over \rho_0}=b^2\mathscr{M}^2.
\end{equation}

Equation~(\ref{conte}) matches the density contrast for the non-magnetic regime \citep[see e.g.,][]{1997PNJ}. This is not surprising, because in this case we are assuming that the gas and the magnetic field are not coupled. Therefore, amplification of the magnetic field with density is not expected under these conditions.

\subsubsection{Second case: $B\propto \rho^{1/2}$}

In the intermediate case in which $B\propto \rho^{1/2}$, we again find a second-order equation for the density contrast, but with a dependence on the magnetic field expressed in terms of $\beta_0$. From Equation (\ref{2-origin}), we obtain

\begin{equation*}%\label{quad}
\left(1+\beta_0^{-1}\right)\left({\rho_2 \over \rho_0}\right)^2-\left(b^2\mathscr{M}^2+1+\beta_0^{-1}\right)\left({\rho_2\over \rho_0}\right)+b^2\mathscr{M}^2=0.
\end{equation*}
This equation has the solution:

\begin{equation}\label{quad}
{\rho_2 \over \rho_0}=b^2\mathscr{M}^2\left({\beta_0 \over \beta_0 +1}\right).
\end{equation}
In other words, the effect of the magnetic field in this case is to reduce the density contrast by a factor $\beta_0/(\beta_0+1)$. We see from this that in the weak field limit where $\beta_0\rightarrow\infty$, we recover the hydrodynamical result, while for strong fields we have a smaller density contrast in the MHD case than in the non-magnetic case. 

\subsubsection{Third case: $B\propto \rho$}

Finally, we investigate the other extreme case, where the magnetic field strength is proportional to the density. In this case, Equation (\ref{2-origin}) results in a third-order equation,

\begin{equation*}%\label{cub}
\beta_0^{-1}\left({\rho_2 \over \rho_0}\right)^3+\left({\rho_2 \over \rho_0}\right)^{2}-\left(b^2\mathscr{M}^2+1+\beta_0^{-1}\right)\left({\rho_2\over \rho_0}\right)+b^2\mathscr{M}^2=0.
\end{equation*}
The solution for the density contrast is

\begin{equation}\label{cub}
{\rho_2 \over \rho_0}={1\over 2}\left(-1-\beta_0+ \sqrt[]{\left(1+\beta_0\right)^2+4b^2\mathscr{M}^2\beta_0}\right).
\end{equation}

\subsection{Density variance--Mach number relation}\label{sigma-s}

In the previous section, we presented three different expressions for the density contrast. They correspond to three different assumptions regarding the relationship $B\propto \rho^\alpha$, with $\alpha=0,1/2$, and 1. We now determine the density variance of a fluid in which there are many shocks, for each of these three cases.

We start by noting that in a highly supersonic flow, the dominant contribution to the integral in Equation (4) will come from shocked regions, and thus we can consider this equation as a volume average over an ensemble of many shocks. We next assume that we can approximate the value of this integral with the result of integrating over a single ``average'' shock of the kind considered in the previous section. As we already know the density contrast of this representative shock, the only thing that remains to be done before we can solve Equation (4) is to determine the appropriate volume over which to integrate.

We approximate the cloud as a cubic box of side L, and consider an infinitesimal part of its volume $\dif V$ that encloses one shock. Therefore, the size of $\dif V$ depends on the size of the shock itself

\begin{equation}\label{difer}
\dif V\approx\dif V_{sh}.
\end{equation}

To define the shock volume, we make use of an approximation introduced by \citet{2011PN}, where the volume of the shock is given by the area of the box face times the shock width $\lambda$, $V_{\rm sh} = L^{2} \lambda$. However, in the absence of viscosity, it is not straightforward to define the shock width $\lambda$. Therefore, we follow \citet{2011PN} and assume that the shock width, if the compression is driven at the box scale, is given by 

\begin{equation}
\lambda \simeq \theta L \rho_{0} / \rho_{2},
\end{equation}
where $\theta$ is the integral scale of the turbulence. Then, the volume of the shock $V_{\rm sh}$ is given by

\begin{equation}\label{vsh}
V_{\rm sh} \simeq \theta L^{3} \frac{\rho_{0}}{\rho_{2}}.
\end{equation}
For turbulence driven on large scales, as appears to be the case in real molecular clouds \citep{2002OssenkopfMacLow,2009Bruntetal}, we have $\theta \simeq 1$. Having made the assumption that the appropriate volume over which to average is the volume of our representative shock, and considering Equation \ref{difer}, we approximate $\dif V$ by

\begin{equation}\label{difer}
\dif V = L^3\left({\rho_0 \over \rho_2}\right)^2 \dif \left({\rho_2\over \rho_0}\right).
\end{equation}

Finally, inserting Equation (\ref{difer}) into Equation (\ref{std}), yields

\begin{equation}\label{resstd}
\sigma_{\rho/\rho_0}^2=\int_1^{\rho \over \rho_0} \left(1-{\rho_0 \over \rho_2}\right)^2 \dif \left({\rho_2\over\rho_0}\right)={\rho \over \rho_0} -{\rho_0 \over \rho} -2\ln {\left({\rho \over \rho_0}\right)}.
\end{equation}
It is important to note that in this formulation, Equation~(\ref{resstd}) is physically meaningless if the lower limit of the integral is set between $0<\rho/\rho_0<1$. It is due to the definition adopted for the shock width (Eq. \ref{vsh}), where the shock thickness is defined only for $\rho_2/\rho_0>1$. For highly supersonic turbulence, which is the regime that concerns us, the assumption $\rho\gg\rho_0$ is valid. Then, the first term in Equation (\ref{resstd}) dominates the variance and we get
\begin{equation}
\sigma_{\rho/\rho_0}^2 \approx {\rho \over \rho_0}.
\end{equation}

For practical reasons, we prefer to consider the variance of the logarithm of the density contrast, $s=\ln(\rho/\rho_0)$, instead of the variance of the linear density when we will compare this analytical model with numerical simulations. These variances are related by \citep[e.g.,][]{2008Federrathetala,2011Priceetal}

\begin{equation}\label{std2}
\sigma_{s}^2=\ln\left[1+\sigma_{\rho/\rho_0}^2\right].
\end{equation}

We now insert the three cases considered in \S\ref{rhocontrast} into Equation (\ref{std2}), in order to obtain the density variance--Mach number relation. The subscripts of the following results are chosen based on the value $\alpha=0$, $1/2$ and 1 of the $B\propto\rho^\alpha$ relationship.

\begin{itemize}
\item {\boldmath$B$} {\bf independent of} {\boldmath$\rho$}

The density variance in this case is exactly the same as for the purely hydrodynamical, isothermal case, 

\begin{equation}\label{HD2}
\sigma_{s,0}^2=\ln\left[1+{b^2\mathscr{M}^2}\right].
\end{equation}

\item {\boldmath$B\propto \rho^{1/2}$}

In this case, the density variance is:

\begin{equation}\label{supreme}
\sigma_{s,1/2}^2=\ln\left[1+b^2\mathscr{M}^2\left({\beta_0 \over \beta _0+1}\right)\right].
\end{equation}

This relation is similar to Equation (\ref{HD2}) except for a correction factor due to the effects of magnetic fields, which is a function of the plasma $\beta_0$ only.

\item {\boldmath{$B\propto \rho$}}

Finally, the density variance--Mach number relation in this case is given by

\begin{equation}\label{supreme2}
\sigma_{s,1}^2=\ln\left[1+{1\over 2}\left(-1-\beta_0+ \sqrt[]{\left(1+\beta_0\right)^2+4b^2\mathscr{M}^2\beta_0}\right)\right].
\end{equation}

The density variance has a strong dependence on $\beta_0$, leaving the rms Mach number as a marginal quantity in this relation.
\end{itemize}

In the last two cases, when $\beta_0\rightarrow 0$, the Alf\'enic velocity is much higher than the sound speed, and both relations approach zero. In this scenario, the magnetic pressure is infinitely large and prevents density fluctuations from forming. The gas is ``frozen'' in the magnetic field. In the opposite limit, when $\beta_0\rightarrow \infty$, Equation (\ref{supreme}) and Equation (\ref{supreme2}) simplify to the purely hydrodynamical case, as expected. In the next section, we are going to test these cases with numerical simulations.

\section{Numerical test of the analytical model}\label{test}

\subsection{Simulations}

We have performed simulations of the evolution of the turbulent, dense, inviscid, magnetised (MHD) and unmagnetised (HD), isothermal interstellar medium using a modified version of the {\sc zeus-mp} hydrodynamical code \citep{2000Norman,2006Hayesetal}. We neglect chemical reactions in order to study the effects of magnetic fields in molecular clouds, leaving the inclusion of the effects of chemistry \citep{2010Gloveretal} for a future study.

Each of our simulations begins with an initially uniform gas distribution, with a mean hydrogen number density of $n_0=1000$ cm$^{-3}$ and a resolution of 256$^3$ cells. The initial velocity field is turbulent, with power concentrated on large scales, between wave numbers $k=1$ and 2 and with an initial rms velocity of $5$ km\,s$^{-1}$. Moreover, we drive the turbulence so as to maintain approximately the same rms velocity throughout the simulations, following the method described in \citet{1998MacLowetal} and \citet{1999MacLow}. We do not perform a Helmholtz decomposition of the force field, and thus the turbulent forcing consists of a natural mixture of solenoidal and compressive modes, i.e., $F_\mathrm{sol}/(F_\mathrm{sol}+F_\mathrm{comp})\approx2/3$. Note that \citet{2008Federrathetala,2010Federrathetal} tested the two limiting cases of purely solenoidal (divergence-free) and purely compressive (curl-free) forcing, as well as various mixtures of solenoidal and compressive modes of the turbulent forcing. They found a strong influence on the density PDF, producing a three times larger standard deviation for compressive forcing compared to solenoidal forcing. They parameterised the influence of the forcing by introducing the $b$-parameter in Equation~(\ref{HD}). Purely solenoidal forcing is characterised by $b=1/3$, while purely compressive forcing gives $b=1$. For the natural mixture, they find $b\approx0.4$. Using the present set of numerical models, we confirm that using $b = 0.4$ for the natural mixture of forcing modes used here gives the best fits with our analytically derived density variance--Mach number relation. The temperature of the gas is constant and fixed to an initial value $T_0=1062$, 170, 42 and 15 K, in order to sample a large set of Mach numbers $\langle\mathscr{M}\rangle\simeq2$, 5, 10 and 17, respectively. We adopt periodic boundary conditions for the gas using a cubical simulation volume with a side length $L=20\,$pc, such that the turbulent crossing time, $T_{\rm cross}=L/(2c_s\mathscr{M})\approx2$ Myr. We present results from $t=3\,T_{\rm cross}\approx 5.7$ Myr, sampled every $0.17\,T_{\rm cross}$, and evolved until $t=4\,T_{\rm cross}\approx 7.6$ Myr. This period of time is long enough to expect the turbulence to have reached a statistical steady state (\citealp*{2009Federrathetal}; \citealp{2010Federrathetal,2010Gloveretal,2010PriceFederrath}). This simulation time might be also short enough to obtain reliable results for the initial phase of star formation, when self-gravity did not yet have a large effect on the dynamics. In order to concentrate on turbulent compression alone, we neglect self-gravity in the present experiments. 

For the MHD cases, the simulations begin with a uniform magnetic field that is initially oriented parallel to the $z$-axis of the simulation. Four of these simulations begin with an initial magnetic field strength $B_i=5.85\,\mu$G, which is our standard magnetic field strength hereafter.  We also perform three MHD runs with $B_i=10$, 20 and 60$\,\mu$G, with $\mathscr{M}=10$, to check the behaviour of the results with increasing magnetic field strengths. We note that as the simulations run, dynamo amplification can lead to increased field strength, and thus we use the instantaneous magnetic field strength to compute $\beta_{0}$. Nevertheless, for simplicity we use the initial value of the magnetic field strength to label runs MHD-B2, MHD-B20 and MHD-B60.

In Table 1, we list the simulations that we have performed. In our labels, we use ``H'' to denote a hydrodynamic run and ``MHD'' to denote a magnetohydrodynamic run. Our multiple runs with fixed (or zero) magnetic field strength but different sound-speeds are labelled with an ``M", followed by the (approximate) rms Mach number of the simulation. Finally, the three runs in which we examined the effect of varying the initial magnetic field strength are labelled with a ``B'', followed by the initial field strength in $\mu$G. In Table 1, we also list the values of the quantities: $\beta_0$, the rms Alfv\'enic Mach number $\mathscr{M}_{A,0}=v_{0}/v_{A,0}$ and the sonic Mach number. They are measured in every cell and then are spatially averaged over the datacube. The brackets denote the time average over the seven snapshots, and the $1\sigma$ shows the temporal standard deviation around the mean values.

\begin{table}
\caption{List of simulations.}
\begin{tabular}{l@{  }c@{ }r@{ }c@{ }l@{ }r@{ }c@{ }l@{ }r@{ }c@{ }l@{ }r@{ }c@{ }l@{ }}
\hline
 & $B_i$ &$\langle\beta_0\rangle$&$\pm$&$1\sigma$& $\langle\mathscr{M}_{A,0}\rangle$& $\pm$ & $1\sigma$&$\langle\sigma_s\rangle$&$\pm$& 1$\sigma$ &$\langle\mathscr{M}\rangle$&$\pm$ & $1\sigma$  \\
 \hline
 HD-M2 & 0 &  &$\infty$& & & 0 & & 0.77 &$\pm$& 0.02  & 2.21 &$\pm$& 0.02  \\
 HD-M5 & 0 &  &$\infty$& & & 0 & & 1.3&$\pm$&0.1 & 5.4 &$\pm$& 0.1  \\
 HD-M10 &  0 & &$\infty$& & & 0 & & 1.7&$\pm$&0.1& 10.6 &$\pm$& 0.2  \\
 HD-M17 &  0 & &$\infty$& & & 0 & &1.92&$\pm$&0.09& 17.6 &$\pm$& 0.5 \\
 MHD-M2 &  5.85 & 25 &$\pm$& 5 & 8.1 &$\pm$& 0.9& 0.69&$\pm$&0.02& 2.09 &$\pm$& 0.02 \\
 MHD-M5 &  5.85 &  4.8 &$\pm$& 0.4 & 8.4 &$\pm$& 0.8& 1.18&$\pm$&0.04& 4.98 &$\pm$& 0.07 \\
 MHD-M10 & 5.85 &  1.4 &$\pm$& 0.5 & 9 &$\pm$& 3 & 1.47&$\pm$&0.06& 10.2 &$\pm$& 0.3\\
 MHD-M17 & 5.85 &   0.3 &$\pm$& 0.1& 7 &$\pm$& 2& 1.61&$\pm$&0.06 & 16.8 &$\pm$& 0.5 \\
 MHD-B2 & 2 &  11.3 &$\pm$& 0.5 & 27 &$\pm$& 2& 1.58&$\pm$&0.09 & 10.5 &$\pm$& 0.2 \\
 MHD-B20 & 20 &  0.083 &$\pm$& 0.005& 1.94 &$\pm$& 0.06& 1.48&$\pm$&0.01 & 9.9 &$\pm$& 0.2 \\
 MHD-B60 & 60 &  0.030 &$\pm$& 0.001& 1.24 &$\pm$& 0.03& 1.34&$\pm$&0.01 & 10.3 &$\pm$& 0.1\\
 
 \hline
 \label{table}
\end{tabular}
 
 $B_i$ -- initial magnetic field strength in $\mu$G. \\ 
 $\beta_0$ -- mean thermal to instantaneous magnetic pressure ratio. \\ 
 $\mathscr{M}_{A,0}$ -- rms Alfv\'enic Mach number. \\ 
 $\sigma_s$ -- density variance. \\
 $\mathscr{M}$ -- rms Mach number. \\
 The brackets indicate the time average calculated over the snapshots after averaging over the spatial coordinates.
\end{table}

\subsection{Statistical Analysis}

In this subsection, we explain the method used to measure the density variance for every snapshot in our simulations using the PDF as a robust statistical tool for this analysis \citep{2011Priceetal}. Then, we parameterise the instantaneous $\beta_0$ in terms of $\mathscr{M}$, in the direction of testing numerically the $\sigma_{s}$--$\mathscr{M}$ relations presented in \S \ref{sigma-s}. Finally, we present the comparison between our analytical model and the simulations.

\subsubsection{Probability Density Function (PDF)}\label{1era}

\begin{figure}
\resizebox{8cm}{!}{\rotatebox{0}{\includegraphics{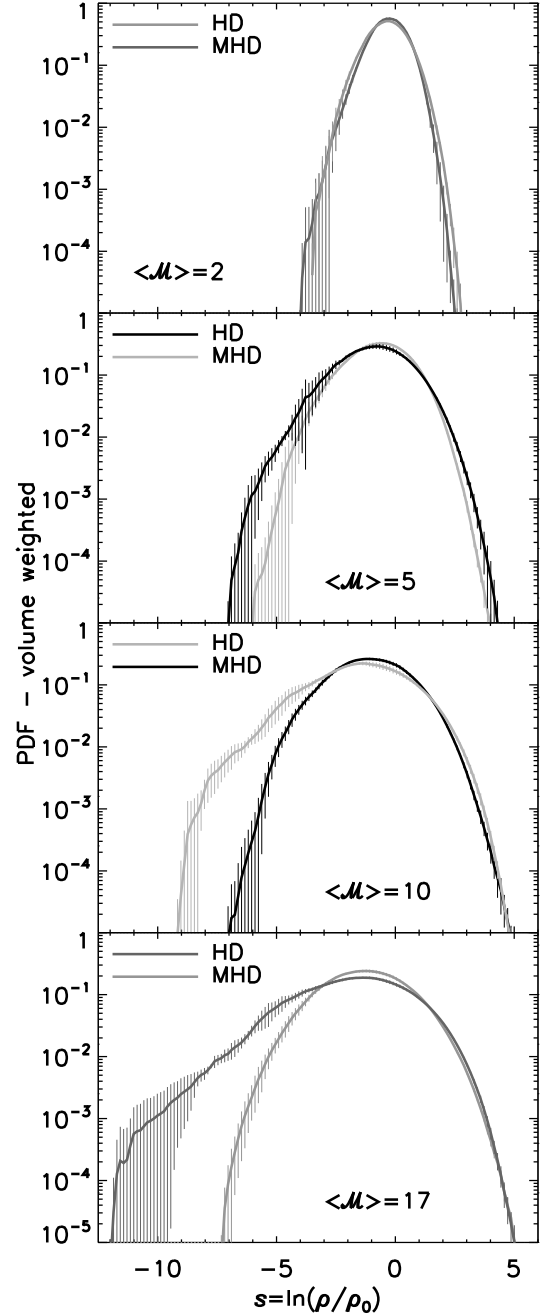}}}
\caption{Dimensionless density PDF for magnetised and unmagnetised molecular clouds with the same initial conditions, $n_0=1000$ cm$^{-3}$, and same turbulent rms velocity, but different sound speed. The most significant features are: 1) the density variance increases with Mach number, and 2) the density variance decreases with magnetic field strength. These simulations have a ratio between thermal pressure and magnetic pressure $\beta_0\lesssim10$. All simulations have a resolution of $256^3$ zones.}
 \label{PDFs}
\end{figure}

In Figure \ref{PDFs}, we plot the volume-weighted dimensionless density PDFs for MHD and HD isothermal gas with the same Mach number for comparison. For these simulations, we find that all the PDFs have a log-normal shape around their peak. However, the PDFs deviate from log-normality especially in the HD simulations at low densities, being more evident for $\mathscr{M} \gtrsim 5$. The error bars in this figure show the 1$\sigma$ variations around the time average. We see that these variations cannot explain the tail at low densities. Therefore, this deviation is not explained by intermittency fluctuations, and deserves further study. However, the low-density tail does not significantly affect our $\sigma_s$ estimates, because the variance is computed from a log-normal fit in a limited interval around the peak, giving the most reliable estimates of $\sigma_s$ \citep[see][]{2011Priceetal}. In this sense, the trend of the time averages observed between MHD and HD simulations shows the magnetic field acting as a density cushion, preventing the gas from reaching very low densities during local expansion. As a consequence, there are larger parts of the volume with density $\rho \approx \rho_0$ in the MHD case than in the HD case.

In order to avoid contamination from intermittency, numerical artefacts, etc., in the wings of the PDFs, we perform a Gaussian fitting only in a data subset selected by $s$, in each simulation. This subset consists of 60\% of the number of bins considered to calculate the density PDF which are distributed symmetrically around the mean, $s_0$. Then, we fit the Gaussian profile given by Equation (\ref{log-norm}) to obtain $\sigma_s$ in every snapshot of the simulations.

\subsubsection{Density variance--rms Mach number test}\label{sigmatest}

In the interest of comparing the density variance--Mach number relation, given by Equation (\ref{supreme}) and Equation (\ref{supreme2}), with the results obtained in the previous subsection, we parameterise the thermal-to-magnetic pressure ratio in terms of the rms Mach number for our sequence of simulations. In this sense, we rewrite Equation (\ref{betafin}) as

\begin{equation}\label{betamach}
\beta_0={2}{\mathscr{M}_{A,0}^2 \over \mathscr{M}^{2}}.
\end{equation} 
Note that this parameter is calculated considering the instantaneous magnetic field strength and not the initial value. 

\begin{figure}
\resizebox{9cm}{!}{\rotatebox{90}{\includegraphics{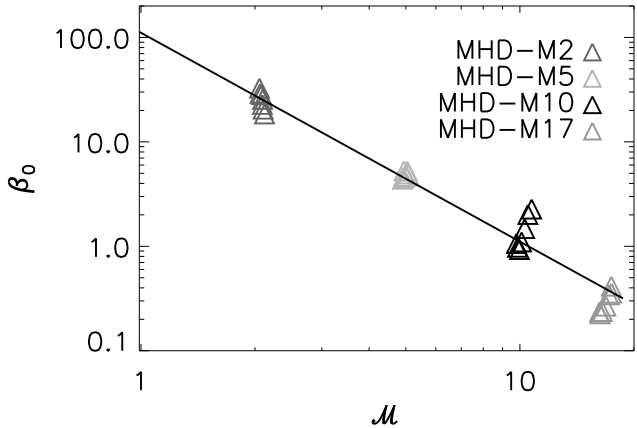}}}
 \caption{Parameterisation of $\beta_0=P_{\rm th}/P_{0,\rm mag}$ with respect to the rms Mach number for the subset of simulations with roughly constant Alfv\'enic Mach number, $\mathscr{M}_{A,0} \approx 8$ (see Table~\ref{table}). The curve is a linear regression of the MHD simulations with $B_i=5.85\mu$G. The linear regression performed to the logarithm of Equation (\ref{betamach}) gives $\beta_0=(111\pm 4)\mathscr{M}^{-2}$.}
 \label{beta}
\end{figure}

Next, we select the four MHD simulations with different rms Mach number, but the same initial magnetic field strength, and use a linear regression considering the logarithm of Equation (\ref{betamach}): $\log_{10}\beta_0=\log_{10}C-2\log_{10}\mathscr{M}$. From the fit shown in Figure~\ref{beta}, we find $C= 111 \pm 4$. In Figure \ref{beta}, we plot $\beta_0$ as a function of the rms Mach number for the different snapshots. The triangles show $\beta_0$ for the selected simulations with $\langle\mathscr{M}\rangle\approx2$, 5, 10 and 17, while the curve shows the linear regression.

\begin{figure*}
\resizebox{18.cm}{!}{\rotatebox{90}{\includegraphics{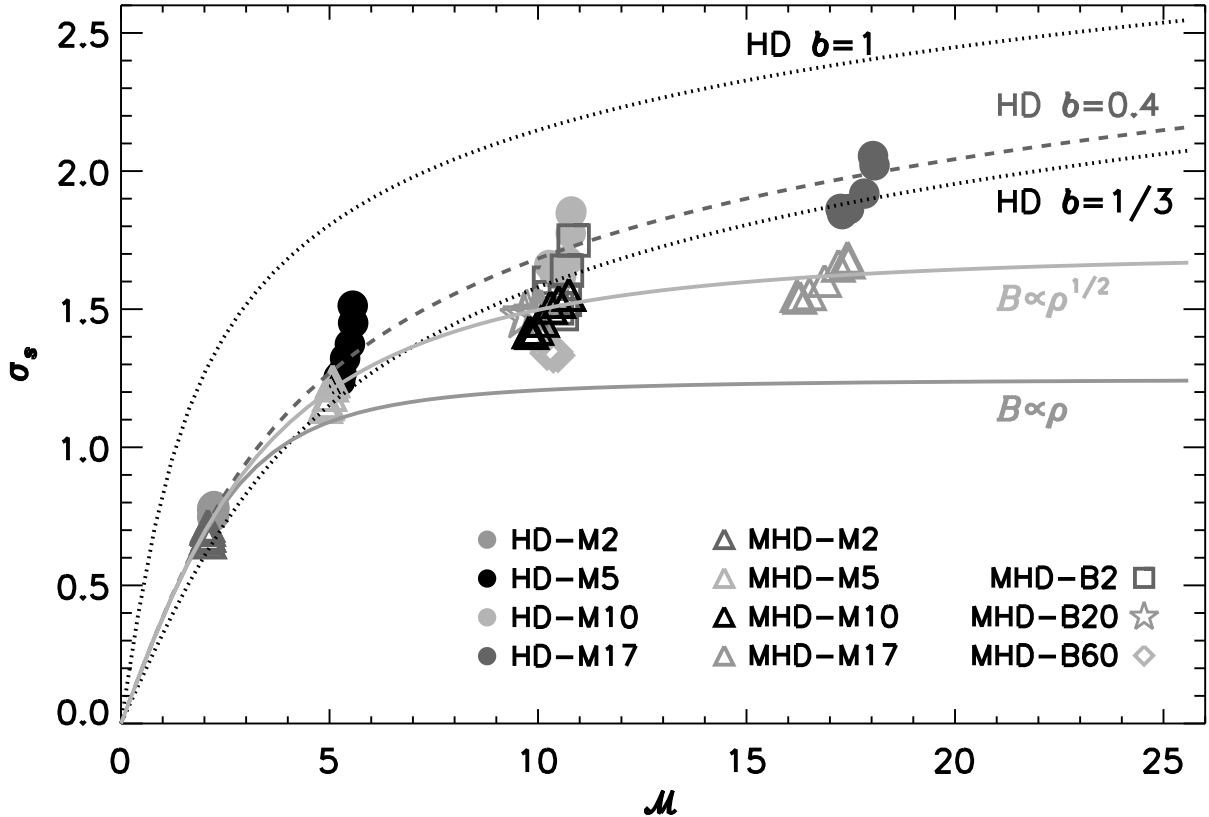}}}
\caption{Standard deviation of the dimensionless density contrast, plotted as a function of the rms Mach number. Circles show the purely hydrodynamical simulations that follow very well the 
\citet{1997PNJ} prediction, $\sigma_{s,HD}^2=\ln(1+b^2\mathscr{M}^2)$, with $b=0.4$, expected for mixed-mode turbulent forcing \citep[][dashed line]{2010Federrathetal}. The dotted lines are for comparison with purely hydrodynamical model, assuming $b=1/3$ for purely solenoidal forcing and $b=1$ for purely compressive forcing 
\citep{2008Federrathetala}. Triangles show the MHD simulations and the two formulas, Eqs. (\ref{supreme}) and (\ref{supreme2}), obtained in this work: $\sigma_{s,1/2}=\{\ln[1+b^2\mathscr{M}^2\beta_0/(\beta_0+1)]\}^{1/2}$ (light grey solid line), and  $\sigma_{s,1}$ (dark grey solid line). Those curves are plotted for $b=0.4$, and using our parameterisation, $\beta_0=(111 \pm 4) \mathscr{M}^{-2}$ from Fig. \ref{beta}. Squares, stars and diamonds show the additional MHD simulations with different rms Alfv\'enic Mach number, $\mathscr{M}_{A,0}\approx 27$ ($B_i=2\,\mu$G), $\mathscr{M}_{A,0}\approx 1.9$ ($B_i=20\,\mu$G), and $\mathscr{M}_{A,0}\approx 1.2$ ($B_i=60\,\mu$G).}
 \label{dinda}
 \end{figure*}

In Figure \ref{dinda}, we combine the dimensionless standard deviation $\sigma_s$, obtained from the fit over the numerical PDFs for every snapshot, and the analytical prediction for the three cases of $B\propto\rho^\alpha$ -- with $\alpha=$0,1/2, and 1 -- as a function of the rms Mach number. For the triangles around a given $\langle\mathscr{M}\rangle$, the HD simulations exhibit larger 
$\sigma_s$ compared with the MHD simulations, as was expected from Figure \ref{PDFs}. For comparison, we plot the analytical prediction given by Equation (\ref{HD2}), $\sigma_{\alpha,0}$. This result matches the prediction provided by \citet{1997PNJ}. However, instead of using their proportionality parameter $b\approx 0.5$, we used the input value $b=0.4$ \citep[][dashed line]{2010Federrathetal}, which is the result of the natural mixing of solenoidal and compressive modes in the turbulent forcing field. We also plot the two extreme cases for the unmagnetised gas, $\sigma_{s,HD}$, with $b=1/3$ (lower dotted line) for purely solenoidal forcing and $b=1$ for purely compressive forcing (upper dotted line) for comparison.

In the same Figure, we superpose Equation (\ref{supreme}, light grey solid line) and Equation (\ref{supreme2}, dark grey solid line), both again with $b=0.4$. We find than the best agreement with the MHD simulations is given by Equation (\ref{supreme}), that is $\sigma_{s,1/2}$. The result obtained for the first case -- $B$ independent of density (Equation \ref{HD2}) -- may account only for low Mach number zones. This case might be appropriate for diffuse clouds \citep{2010Crutcheretal}, where the mean sound speed of the cloud may be of the same order as the rms velocity. Here, at $\mathscr{M} \sim1$, all the three cases converge to the HD result.

Our results are qualitatively in agreement with \citet{2001Ostrikeretal} and \citet{2011Priceetal}. These authors find that the density variance in magnetised gas is significantly lower than in the HD counterparts for simulations with a Mach number $\mathscr{M}\gtrsim 10$. In addition, \citet{2003ChoLazarian} study the density contrast resulting from the Alfv\'enic waves, slow and fast magneto-sonic waves originating in different environments. The authors concluded that the three kinds of waves can coexist in those environments. In the regime that concerns us, $\beta_0\approx 1$ and $5\lesssim\mathscr{M}\lesssim10$, their density contrasts closely match ours.

\begin{figure*}
\resizebox{18.cm}{!}{\includegraphics{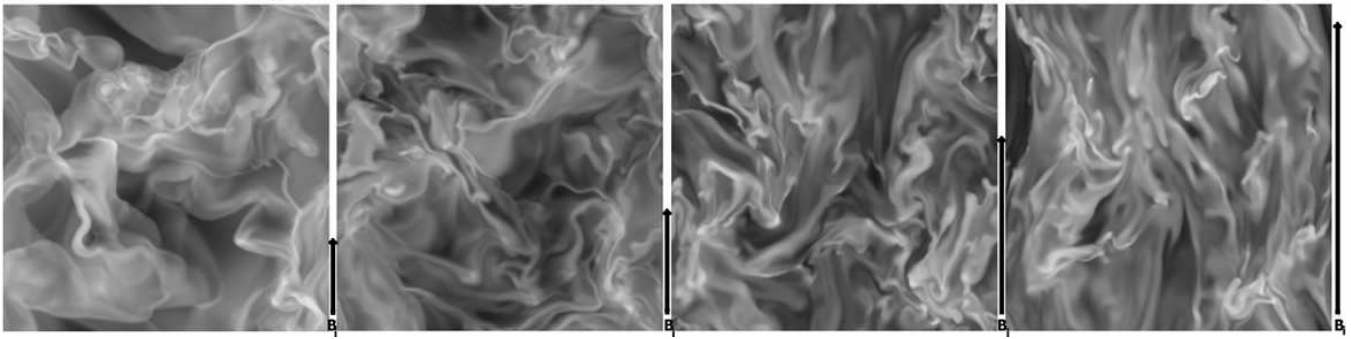}}
\caption{Density slices of the simulations at $t=6$ Myr. The mean magnetic field is oriented along the vertical axis. From left to right: initial magnetic field strength $B_i=2$, 5.85, 20 and 60 $\mu$G. The turbulence remains isotropic for super-Alfv\'enic gas $\mathscr{M}_{A,0}\gg1$, but when it becomes trans-Alfv\'enic or sub Alfv\'enic ($\mathscr{M}_{A,0}\lesssim3$), the turbulence becomes highly anisotropic.}
 \label{dibujito}
 \end{figure*}

\begin{figure}
\resizebox{9cm}{!}{\rotatebox{90}{\includegraphics{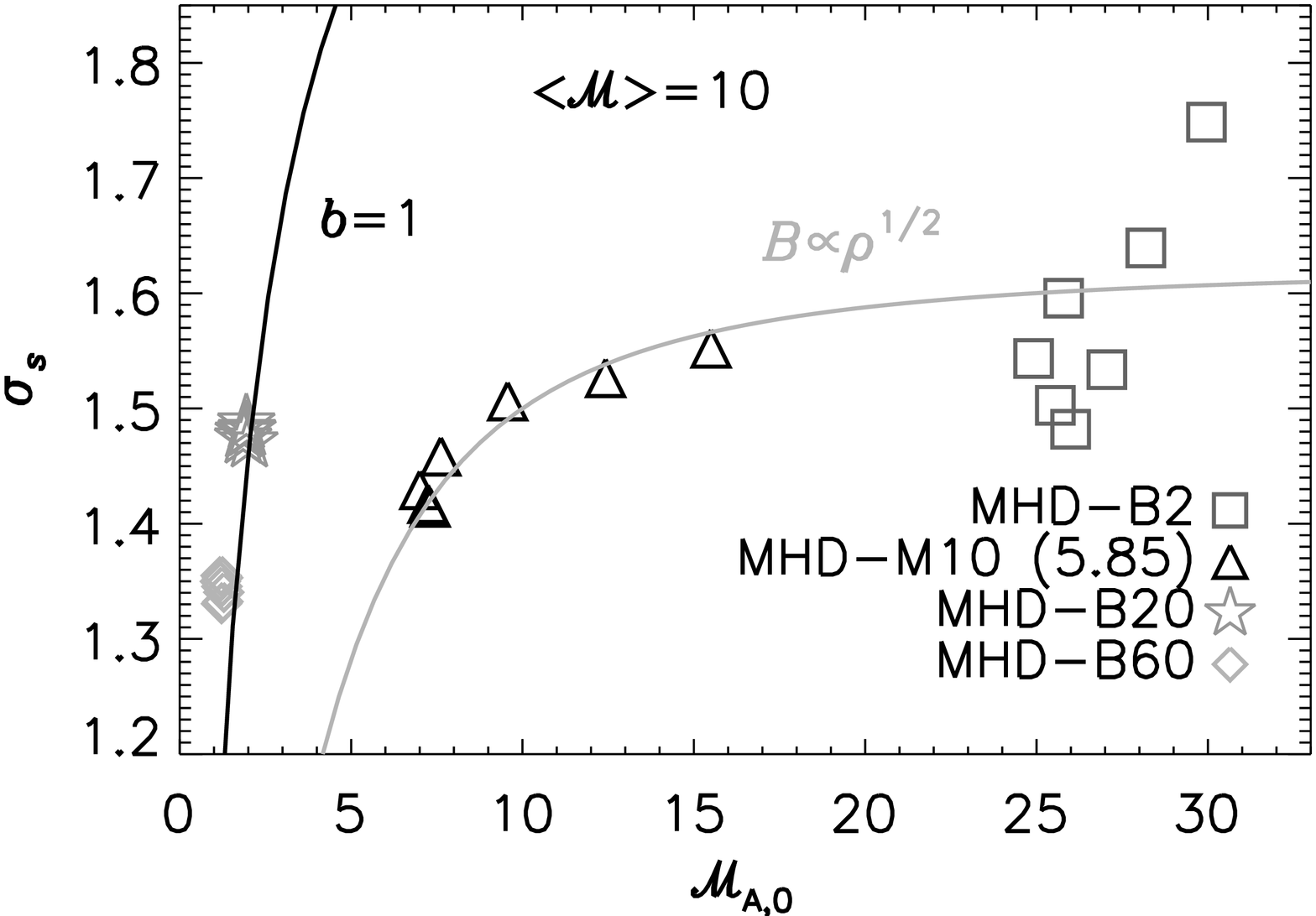}}}
\caption{Standard deviation of the dimensionless density contrast, plotted as a function of the instantaneous rms Alfv\'enic Mach number at $\langle\mathscr{M}\rangle\approx 10$. The different symbols show snapshots of simulations with $\mathscr{M}_{A,0}$ time averages: $\langle \mathscr{M}_{A,0}\rangle\approx 27$ (squares),  $\langle\mathscr{M}_{A,0}\rangle\approx 9$ (triangles),  $\langle\mathscr{M}_{A,0}\rangle\approx 1.9$ (stars), and $\langle\mathscr{M}_{A,0}\rangle\approx 1.2$ (diamonds). When the turbulence becomes trans-Alfv\'enic or sub-Alfv\'enic, $\langle \mathscr{M}_{A,0}\rangle\lesssim 2$ (stars and diamonds), anisotropies arise in the gas, because the back reaction of the magnetic field onto the flow is extremely strong for flows perpendicular to the magnetic field lines.
The grey curve shows our prediction$^1$ using $b\approx 0.4$ that fits very well the data. Meanwhile, the black curve shows our prediction$^1$ considering $b=1$ (corresponding to purely compressive forcing). Although our turbulent forcing in the simulations is by definition mixed, and thus we expect $b\approx 0.4$ \citep{2010Federrathetal}, it is noteworthy to say that $b=1$ gives a good fit to the data with very low $\langle\mathscr{M}_{A,0}\rangle\lesssim 2$.}
 \label{sig-alf}
 \end{figure}

To test the validity of our results for different Alfv\'enic Mach numbers, we also performed three simulations with an initial magnetic field strength different from the standard one, with $\mathscr{M}_{A,0}\approx27$, 1.9, and 1.2, at $\langle\mathscr{M}\rangle\approx 10$ (empty squares in Figure \ref{dinda}). Our model works well for $\mathscr{M}_{A,0}\gtrsim 6$, but breaks down for our test with $\mathscr{M}_{A,0}\lesssim 2$. The break occurs when the turbulence becomes trans-Alfv\'enic or sub-Alfv\'enic, i.e., when $\mathscr{M}_{A,0}\lesssim2$. This is due to anisotropies arising in this case, i.e., the turbulence is no longer isotropic, as can be seen in Figure \ref{dibujito}. This is because the back reaction of the magnetic field onto the flow is extremely strong for flows perpendicular to the magnetic field lines, if the turbulence is trans-Alfv\'enic or sub-Alfv\'enic \citep*[see e.g.,][]{2003ChoLazarian, 2010Bruntetal, 2011EsquivelLazarian}. Since our analytic derivation is based on an ensemble average (Eq.~\ref{std}), assuming statistical isotropy, the anisotropies are the most likely cause for the limitation of our model to super-Alfv\'enic turbulence. In Figure~\ref{sig-alf}, we show our prediction (Eq. \ref{supreme})\footnote{Equation (\ref{supreme}) has been written in terms of the instantaneous Alfv\'enic Mach number (Eq. \ref{betamach}), yielding the relation for the density variance: $\sigma_{s,1/2}^2=\ln[1+2b^2\mathscr{M}^2\mathscr{M}_{A,0}^2/(2\mathscr{M}_{A,0}^2+\mathscr{M}^2)]$.} for a fixed Mach number $\mathscr{M}\approx10$ and forcing parameter $b\approx 0.4$, which fits very well the data with $\mathscr{M}_{A,0}\gtrsim 6$. These simulations show high dispersion -- around the time average -- in the density variance and the rms Alfv\'enic Mach number showing the fluctuations of the gas caused by the turbulence dominating the dynamics of the flow, in contraposition of the simulations with small Alfv\'enic Mach number. In the same Figure, we also plot the model curve Eq. (\ref{supreme}) for the same sonic Mach number 10 and $b=1$. Although our turbulent forcing in the simulations is by definition mixed, and thus we expect $b\approx 0.4$ \citep{2010Federrathetal}, we find it interesting to note that $b=1$ -- corresponding to purely compressive forcing -- gives a good fit to the data with very low Alfv\'enic Mach number, $\mathscr{M}_{A,0} \lesssim 2$. We speculate that the density field for very high magnetic field strengths and thus very low Alfv\'enic Mach number starts behaving as if it was driven by purely compressive forcing. This is very different from the compression obtained with solenoidal or mixed forcing, but more similar to compressive forcing, which also directly compresses the gas \citep{2008Federrathetala}. More data at $\mathscr{M}_{A,0} \lesssim 2$ would be needed to sample this region and the transition from $b=0.4$ to 1 in detail, and we just note here that $b=1$ seems to provide a good fit for $\mathscr{M}_{A,0} \lesssim 2$, given the data at hand.

\section{Conclusions}

We presented an analytical prediction for the density variance--Mach number relation in magnetised supersonic turbulent gas. In this formulation, we considered three different cases for the relation between the magnetic field strength and density. The first case assumes that $B$ is independent of $\rho$, the second assumes that $B \propto \rho^{1/2}$, while the third is given by $B\propto \rho$. The three resulting $\sigma_s$--$\mathscr{M}$ relations were tested against numerical simulations. From this analysis we conclude that:  

\begin{itemize}

\item If $B$ is independent of the density, we recover the hydrodynamical prediction of \citet{1997PNJ}. In this case, the gas and the magnetic field are not coupled. Therefore, an amplification of the magnetic field with the shock is not expected. Observationally, \citet{1999Crutcher} found that the magnetic field was independent of the density for diffuse clouds, corresponding to low rms Mach numbers, $\mathscr{M}\lesssim 2$. In this regime, all our predictions converge to the purely hydrodynamical $\sigma_s$--$\mathscr{M}$ relation.

\item For the second case, $B\propto \rho^{1/2}$, we found a one-to-one relation between $\mathscr{M}$, $\beta_0$ and the density variance. This $\sigma_s$--$\mathscr{M}$ relation (Eq.~\ref{supreme}) matches very well our numerical test considering $b=0.4$, which is the input for the natural mixture of compressive-to-solenoidal modes in the turbulent forcing field. This result is in agreement with the ones presented by \citet{2001Ostrikeretal} and  \citet{2011Priceetal}, where they found lower $\sigma_s$ than in the unmagnetised case for $\mathscr{M}\gtrsim 10$. Moreover, \citet{2003ChoLazarian} presented a density contrast that closely matches our result for $\beta_0\approx 1$ and $5\lesssim\mathscr{M}\lesssim10$.

\item For the last case, $B\propto \rho$, the $\sigma_s$--$\mathscr{M}$ relation (Eq.~\ref{supreme2}) predicts a lower density variance than measured in our numerical simulations for $\mathscr{M}\geq 5$, because our simulations are closer to $B\propto\rho^{1/2}$. However, the relation given by Equation~(\ref{supreme2}) would fit better, if $B\propto\rho$.

\item The $\sigma_s$--$\mathscr{M}$ relation obtained for $B\propto \rho^{1/2}$ works very well for intermediate to high Alfv\'enic Mach number, $\mathscr{M}_{A,0}\gtrsim 6$, but breaks down for $\mathscr{M}_{A,0} \lesssim 2$ at $\langle\mathscr{M}\rangle \approx 10$. This probably occurs because in the presence of strong magnetic fields, the turbulence is no longer isotropic. This is because the back reaction of the magnetic field onto the flow is very strong for flows perpendicular to the magnetic field lines.

\end{itemize}

Magnetic fields act as a density cushion in turbulent gas, preventing the gas from reaching very low densities as well as very high densities. We conclude that magnetic fields are an important mechanism for shaping the density variance--Mach number relation, and therefore will change the quantitative predictions in models of the star formation rate, initial mass function or core mass function that depend on these quantities \citep[e.g.][]{2005KrumholzMckee,2011PN,2002PN,2008HennebelleChabrier,2009HennebelleChabrier}. 
\section*{Acknowledgments}

We thanks the anonymous referee for the useful comments that helped to improve this manuscript. F.~Z.~M. thanks Paola Pinilla and Joe Ramsey for reading the manuscript and providing useful comments, as well as the International Max Planck Research School for Astronomy and Cosmic Physics at the University of Heidelberg, and the Heidelberg Graduate School of Fundamental Physics. F.~Z.~M.~received funding from the German Bundesministerium f\"ur Bildung und Forschung for funding via the ASTRONET project STAR FORMAT (grant 05A09VHA) C.F.~received funding from a Discovery Projects fellowship by the Australian Research Council (grant~DP110102191) and from the European Research Council (FP7/2007-2013 Grant Agreement no.~247060). All authors acknowledge subsidies from the Baden-W\"urttemberg-Stiftung for contract research via grant  P-LS-SPII/18, and the Deutsche Forschungsgemeinschaft via SFB project 881 ``The Milky Way System'' (subprojects B1, B2, B3, and B5) as well as the SPP 1573 ``The Physics of the ISM''. The simulations described in this paper were performed using the {\em Ranger} cluster at the Texas Advanced Computing Center, using time allocated as part of Teragrid project TG-MCA99S024.

\bibliography{pp}

\begin{thebibliography}{63}
\expandafter\ifx\csname natexlab\endcsname\relax\def\natexlab#1{#1}\fi

\bibitem[{{Banerjee} {et~al}\mbox{.}(2009){Banerjee}, {V{\'a}zquez-Semadeni},
  {Hennebelle}, \& {Klessen}}]{2009Banerjeeetal}
{Banerjee} R., {V{\'a}zquez-Semadeni} E., {Hennebelle} P., {Klessen} R.~S.,
  2009, \mnras, 398, 1082

\bibitem[{{Bertram} {et~al}\mbox{.}(2012){Bertram}, {Federrath}, {Banerjee}, \&
  {Klessen}}]{2012Bertrametal}
{Bertram} E., {Federrath} C., {Banerjee} R., {Klessen} R.~S., 2012, \mnras,
  420, 3163

\bibitem[{{Brunt} {et~al}\mbox{.}(2010){Brunt}, {Federrath}, \&
  {Price}}]{2010Bruntetal}
{Brunt} C.~M., {Federrath} C., {Price} D.~J., 2010, \mnras, 403, 1507

\bibitem[{{Brunt} {et~al}\mbox{.}(2009){Brunt}, {Heyer}, \& {Mac
  Low}}]{2009Bruntetal}
{Brunt} C.~M., {Heyer} M.~H., {Mac Low} M.-M., 2009, \aap, 504, 883

\bibitem[{{Cho} \& {Lazarian}(2003)}]{2003ChoLazarian}
{Cho} J., {Lazarian} A., 2003, \mnras, 345, 325

\bibitem[{{Cho} \& {Kim}(2011)}]{2011ChoKim}
{Cho} W., {Kim} J., 2011, \mnras, 410, L8

\bibitem[{{Collins} {et~al}\mbox{.}(2011){Collins}, {Padoan}, {Norman}, \&
  {Xu}}]{2011Collinsetal}
{Collins} D.~C., {Padoan} P., {Norman} M.~L., {Xu} H., 2011, \apj, 731, 59

\bibitem[{{Crutcher} {et~al}\mbox{.}(2003){Crutcher}, {Heiles}, \&
  {Troland}}]{2003Crutcheretal}
{Crutcher} R., {Heiles} C., {Troland} T., 2003, in Lecture Notes in Physics,
  Berlin Springer Verlag, Vol. 614, Turbulence and Magnetic Fields in
  Astrophysics, {E.~Falgarone \& T.~Passot}, ed., pp. 155--181

\bibitem[{{Crutcher}(1999)}]{1999Crutcher}
{Crutcher} R.~M., 1999, \apj, 520, 706

\bibitem[{{Crutcher} {et~al}\mbox{.}(2009){Crutcher}, {Hakobian}, \&
  {Troland}}]{2009Crutcheretal}
{Crutcher} R.~M., {Hakobian} N., {Troland} T.~H., 2009, \apj, 692, 844

\bibitem[{{Crutcher} {et~al}\mbox{.}(2010){Crutcher}, {Wandelt}, {Heiles},
  {Falgarone}, \& {Troland}}]{2010Crutcheretal}
{Crutcher} R.~M., {Wandelt} B., {Heiles} C., {Falgarone} E., {Troland} T.~H.,
  2010, \apj, 725, 466

\bibitem[{{Dyson} \& {Williams}(1980)}]{1980DysonWilliams}
{Dyson} J.~E., {Williams} D.~A., 1980, Physics of the interstellar medium, New
  York, Halsted Press, 1980.~204 p., {{Dyson}, J.~E.~\& {Williams}, D.~A.}, ed.

\bibitem[{{Elmegreen} \& {Scalo}(2004)}]{2004Elmegreen}
{Elmegreen} B.~G., {Scalo} J., 2004, \araa, 42, 211

\bibitem[{{Esquivel} \& {Lazarian}(2011)}]{2011EsquivelLazarian}
{Esquivel} A., {Lazarian} A., 2011, \apj, 740, 117

\bibitem[{{Federrath} {et~al}\mbox{.}(2008{\natexlab{a}}){Federrath}, {Glover},
  {Klessen}, \& {Schmidt}}]{2008Federrathetalb}
{Federrath} C., {Glover} S.~C.~O., {Klessen} R.~S., {Schmidt} W.,
  2008{\natexlab{a}}, Physica Scripta Volume T, 132, 014025

\bibitem[{{Federrath} {et~al}\mbox{.}(2008{\natexlab{b}}){Federrath},
  {Klessen}, \& {Schmidt}}]{2008Federrathetala}
{Federrath} C., {Klessen} R.~S., {Schmidt} W., 2008{\natexlab{b}}, \apjl, 688,
  L79

\bibitem[{{Federrath} {et~al}\mbox{.}(2009){Federrath}, {Klessen}, \&
  {Schmidt}}]{2009Federrathetal}
{Federrath} C., {Klessen} R.~S., {Schmidt} W., 2009, \apj, 692, 364

\bibitem[{{Federrath} {et~al}\mbox{.}(2010){Federrath}, {Roman-Duval},
  {Klessen}, {Schmidt}, \& {Mac Low}}]{2010Federrathetal}
{Federrath} C., {Roman-Duval} J., {Klessen} R.~S., {Schmidt} W., {Mac Low}
  M.-M., 2010, \aap, 512, A81

\bibitem[{{Federrath} {et~al}\mbox{.}(2011){Federrath}, {Sur}, {Schleicher},
  {Banerjee}, \& {Klessen}}]{2011Federrathetal}
{Federrath} C., {Sur} S., {Schleicher} D.~R.~G., {Banerjee} R., {Klessen}
  R.~S., 2011, \apj, 731, 62

\bibitem[{{Fleck}(1982)}]{1982Fleck}
{Fleck}, Jr. R.~C., 1982, \mnras, 201, 551

\bibitem[{{Glover} {et~al}\mbox{.}(2010){Glover}, {Federrath}, {Mac Low}, \&
  {Klessen}}]{2010Gloveretal}
{Glover} S.~C.~O., {Federrath} C., {Mac Low} M.-M., {Klessen} R.~S., 2010,
  \mnras, 404, 2

\bibitem[{{Glover} \& {Mac Low}(2007)}]{2007GloveryMacLow}
{Glover} S.~C.~O., {Mac Low} M.-M., 2007, \apj, 659, 1317

\bibitem[{{Hayes} {et~al}\mbox{.}(2006){Hayes}, {Norman}, {Fiedler}, {Bordner},
  {Li}, {Clark}, {ud-Doula}, \& {Mac Low}}]{2006Hayesetal}
{Hayes} J.~C., {Norman} M.~L., {Fiedler} R.~A., {Bordner} J.~O., {Li} P.~S.,
  {Clark} S.~E., {ud-Doula} A., {Mac Low} M.-M., 2006, \apjs, 165, 188

\bibitem[{{Hennebelle} \& {Chabrier}(2008)}]{2008HennebelleChabrier}
{Hennebelle} P., {Chabrier} G., 2008, \apj, 684, 395

\bibitem[{{Hennebelle} \& {Chabrier}(2009)}]{2009HennebelleChabrier}
{Hennebelle} P., {Chabrier} G., 2009, \apj, 702, 1428

\bibitem[{{Hennebelle} \& {P{\'e}rault}(2000)}]{2000HennebellePerault}
{Hennebelle} P., {P{\'e}rault} M., 2000, \aap, 359, 1124

\bibitem[{{Kainulainen} {et~al}\mbox{.}(2009){Kainulainen}, {Beuther},
  {Henning}, \& {Plume}}]{2009Kainulainenetal}
{Kainulainen} J., {Beuther} H., {Henning} T., {Plume} R., 2009, \aap, 508, L35

\bibitem[{{Kim} {et~al}\mbox{.}(2001){Kim}, {Balsara}, \& {Mac
  Low}}]{2001Kimetal}
{Kim} J., {Balsara} D., {Mac Low} M.-M., 2001, Journal of Korean Astronomical
  Society, 34, 333

\bibitem[{{Klessen}(2000)}]{2000Klessen}
{Klessen} R.~S., 2000, \apj, 535, 869

\bibitem[{{Klessen} \& {Burkert}(2000)}]{2000KlessenBurkert}
{Klessen} R.~S., {Burkert} A., 2000, \apjs, 128, 287

\bibitem[{{Kritsuk} {et~al}\mbox{.}(2007){Kritsuk}, {Norman}, {Padoan}, \&
  {Wagner}}]{2007Kritsuketal}
{Kritsuk} A.~G., {Norman} M.~L., {Padoan} P., {Wagner} R., 2007, \apj, 665, 416

\bibitem[{{Kritsuk} {et~al}\mbox{.}(2011){Kritsuk}, {Norman}, \&
  {Wagner}}]{2011Kritsuketal}
{Kritsuk} A.~G., {Norman} M.~L., {Wagner} R., 2011, \apjl, 727, L20

\bibitem[{{Krumholz} \& {McKee}(2005)}]{2005KrumholzMckee}
{Krumholz} M.~R., {McKee} C.~F., 2005, \apj, 630, 250

\bibitem[{{Landau} \& {Lifshitz}(1987)}]{1987LandauLifshitz}
{Landau} L.~D., {Lifshitz} E.~M., 1987, {Fluid mechanics}, {Landau, L.~D.~\&
  Lifshitz, E.~M.}, ed.

\bibitem[{{Lemaster} \& {Stone}(2008)}]{2008Lemaster}
{Lemaster} M.~N., {Stone} J.~M., 2008, \apjl, 682, L97

\bibitem[{{Lequeux}(2005)}]{2005Lequeux}
{Lequeux} J., 2005, The interstellar medium, Translation from the French
  language edition of: Le Milieu Interstellaire by James Lequeux, EDP Sciences,
  2003 Edited by J.~Lequeux.~ Astronomy and astrophysics library, Berlin:
  Springer, 2005, {Lequeux, J.}, ed.

\bibitem[{{Li} {et~al}\mbox{.}(2004){Li}, {Norman}, {Mac Low}, \&
  {Heitsch}}]{2004Lietal}
{Li} P.~S., {Norman} M.~L., {Mac Low} M.-M., {Heitsch} F., 2004, \apj, 605, 800

\bibitem[{{Li} {et~al}\mbox{.}(2003){Li}, {Klessen}, \& {Mac Low}}]{2003Lietal}
{Li} Y., {Klessen} R.~S., {Mac Low} M.-M., 2003, \apj, 592, 975

\bibitem[{{Mac Low}(1999)}]{1999MacLow}
{Mac Low} M.-M., 1999, \apj, 524, 169

\bibitem[{{Mac Low} \& {Klessen}(2004)}]{2004MacLowKlessen}
{Mac Low} M.-M., {Klessen} R.~S., 2004, Reviews of Modern Physics, 76, 125

\bibitem[{{Mac Low} {et~al}\mbox{.}(1998){Mac Low}, {Klessen}, {Burkert}, \&
  {Smith}}]{1998MacLowetal}
{Mac Low} M.-M., {Klessen} R.~S., {Burkert} A., {Smith} M.~D., 1998, Physical
  Review Letters, 80, 2754

\bibitem[{{McKee} \& {Ostriker}(2007)}]{2007McKeeOstriker}
{McKee} C.~F., {Ostriker} E.~C., 2007, \araa, 45, 565

\bibitem[{{Mouschovias} \& {Ciolek}(1999)}]{1999MouschoviasCiolek}
{Mouschovias} T.~C., {Ciolek} G.~E., 1999, in NATO ASIC Proc. 540: The Origin
  of Stars and Planetary Systems, {C.~J.~Lada \& N.~D.~Kylafis}, ed., p. 305

\bibitem[{{Nordlund} \& {Padoan}(1999)}]{1999NP}
{Nordlund} {\AA}.~K., {Padoan} P., 1999, in Interstellar Turbulence, {J.~Franco
  \& A.~Carraminana}, ed., p. 218

\bibitem[{{Norman}(2000)}]{2000Norman}
{Norman} M.~L., 2000, in Revista Mexicana de Astronomia y Astrofisica, vol. 27,
  Vol.~9, Revista Mexicana de Astronomia y Astrofisica Conference Series,
  {S.~J.~Arthur, N.~S.~Brickhouse, \& J.~Franco}, ed., pp. 66--71

\bibitem[{{Ossenkopf} \& {Mac Low}(2002)}]{2002OssenkopfMacLow}
{Ossenkopf} V., {Mac Low} M.-M., 2002, \aap, 390, 307

\bibitem[{{Ostriker} {et~al}\mbox{.}(2001){Ostriker}, {Stone}, \&
  {Gammie}}]{2001Ostrikeretal}
{Ostriker} E.~C., {Stone} J.~M., {Gammie} C.~F., 2001, \apj, 546, 980

\bibitem[{{Padoan} \& {Nordlund}(1999)}]{1999PN}
{Padoan} P., {Nordlund} {\AA}., 1999, \apj, 526, 279

\bibitem[{{Padoan} \& {Nordlund}(2002)}]{2002PN}
{Padoan} P., {Nordlund} {\AA}., 2002, \apj, 576, 870

\bibitem[{{Padoan} \& {Nordlund}(2011)}]{2011PN}
{Padoan} P., {Nordlund} {\AA}., 2011, \apj, 730, 40

\bibitem[{{Padoan} {et~al}\mbox{.}(1997){Padoan}, {Nordlund}, \&
  {Jones}}]{1997PNJ}
{Padoan} P., {Nordlund} A., {Jones} B.~J.~T., 1997, \mnras, 288, 145

\bibitem[{{Passot} \& {V{\'a}zquez-Semadeni}(1998)}]{1998PassotyVS}
{Passot} T., {V{\'a}zquez-Semadeni} E., 1998, \pre, 58, 4501

\bibitem[{{Pope} \& {Ching}(1993)}]{1993PopeChing}
{Pope} S.~B., {Ching} E. S.~C., 1993, \phfla, 5, 1529

\bibitem[{{Press} \& {Schechter}(1974)}]{1974PressSchechter}
{Press} W.~H., {Schechter} P., 1974, \apj, 187, 425

\bibitem[{{Price} \& {Federrath}(2010)}]{2010PriceFederrath}
{Price} D.~J., {Federrath} C., 2010, \mnras, 406, 1659

\bibitem[{{Price} {et~al}\mbox{.}(2011){Price}, {Federrath}, \&
  {Brunt}}]{2011Priceetal}
{Price} D.~J., {Federrath} C., {Brunt} C.~M., 2011, \apjl, 727, L21

\bibitem[{{Scalo} {et~al}\mbox{.}(1998){Scalo}, {Vazquez-Semadeni}, {Chappell},
  \& {Passot}}]{1998Scaloetal}
{Scalo} J., {Vazquez-Semadeni} E., {Chappell} D., {Passot} T., 1998, \apj, 504,
  835

\bibitem[{{Schleicher} {et~al}\mbox{.}(2010){Schleicher}, {Banerjee}, {Sur},
  {Arshakian}, {Klessen}, {Beck}, \& {Spaans}}]{2010Schleicheretal}
{Schleicher} D.~R.~G., {Banerjee} R., {Sur} S., {Arshakian} T.~G., {Klessen}
  R.~S., {Beck} R., {Spaans} M., 2010, \aap, 522, A115

\bibitem[{{Schmidt} {et~al}\mbox{.}(2009){Schmidt}, {Federrath}, {Hupp},
  {Kern}, \& {Niemeyer}}]{2009Schmidtetal}
{Schmidt} W., {Federrath} C., {Hupp} M., {Kern} S., {Niemeyer} J.~C., 2009,
  \aap, 494, 127

\bibitem[{{Sur} {et~al}\mbox{.}(2010){Sur}, {Schleicher}, {Banerjee},
  {Federrath}, \& {Klessen}}]{2010Suretal}
{Sur} S., {Schleicher} D.~R.~G., {Banerjee} R., {Federrath} C., {Klessen}
  R.~S., 2010, \apjl, 721, L134

\bibitem[{{V\'azquez-Semadeni}(1994)}]{1994Vazquez-Semadeni}
{V\'azquez-Semadeni} E., 1994, \apj, 423, 681

\bibitem[{{Wada}(2001)}]{2001Wada}
{Wada} K., 2001, \apjl, 559, L41

\bibitem[{{Zinnecker}(1984)}]{1984Zinnecker}
{Zinnecker} H., 1984, \mnras, 210, 43

\end{thebibliography}

\bsp
\label{lastpage}

\end{document}